\documentclass[prb,showpacs,floatfix,amsmath,amssymb,superscriptaddress]{revtex4-1}
\usepackage{amsfonts}
\usepackage{stmaryrd}
\usepackage{bbm}
\usepackage{mathrsfs}
\usepackage{tipa}
\usepackage{amssymb}
\usepackage{txfonts}
\usepackage{graphicx}
\usepackage{dcolumn}
\usepackage{epstopdf}
\usepackage[colorlinks,linkcolor=blue,urlcolor=blue,citecolor=blue]{hyperref}
\usepackage{multirow}
\usepackage{subfigure}
\usepackage{url}
\usepackage{upgreek}

\begin{document}
\newcommand*{\cm}{cm$^{-1}$\,}
\newcommand*{\Tc}{T$_c$\,}
\newcommand{\degree}{^\circ}

\title{Photo-induced new metastable state with modulated Josephson coupling strengths in Pr$_{0.88}$LaCe$_{0.12}$CuO$_4$}%force line break\\

\author{S. J. Zhang}
\affiliation{International Center for Quantum Materials, School of Physics, Peking University, Beijing 100871, China}

\author{Z. X. Wang}
\affiliation{International Center for Quantum Materials, School of Physics, Peking University, Beijing 100871, China}

\author{D. Wu}
\affiliation{International Center for Quantum Materials, School of Physics, Peking University, Beijing 100871, China}

\author{Q. M. Liu}
\affiliation{International Center for Quantum Materials, School of Physics, Peking University, Beijing 100871, China}

\author{L. Y. Shi}
\affiliation{International Center for Quantum Materials, School of Physics, Peking University, Beijing 100871, China}

\author{T. Lin}
\affiliation{International Center for Quantum Materials, School of Physics, Peking University, Beijing 100871, China}

\author{S. L. Li}
\affiliation{Institute of physics,Chinese academy of Sciences, Beijing 100190, China}
\affiliation{Collaborative Innovation Center of Quantum Matter, Beijing, China}

\author{P. C. Dai}
\affiliation{Department of Physics, Rice University, USA}

\author{T. Dong}
\affiliation{International Center for Quantum Materials, School of Physics, Peking University, Beijing 100871, China}

\author{N. L. Wang}
\email{nlwang@pku.edu.cn}
\affiliation{International Center for Quantum Materials, School of Physics, Peking University, Beijing 100871, China}
\affiliation{Collaborative Innovation Center of Quantum Matter, Beijing, China}

\begin{abstract}
Photoexcitations on a superconductor using ultrafast nir-infrared (NIR) pulses, whose energy is much higher than the superconducting energy gap, are expected to suppress/destroy superconductivity by breaking Cooper pairs and excite quasiparticles from occupied state to unoccupied state far above the Fermi level. This appears to be true only for small pumping fluence. Here we show that the intense NIR pumping has different effect. We perform an intense NIR pump, c-axis terahertz probe measurement on an electron-doped cuprate superconductor Pr$_{0.88}$LaCe$_{0.12}$CuO$_4$ with T$_c$=22 K. The measurement indicates that, instead of destroying superconductivity or exciting quasiparticles, the intense NIR pump drives the system from an equilibrium superconducting state with uniform Josephson coupling strength to a new metastable superconducting phase with modulated Josephson coupling strengths below T$_c$.
\end{abstract}

\maketitle
\begin{figure*}
  \centering
  % Requires \usepackage{graphicx}
  \includegraphics[width=12cm]{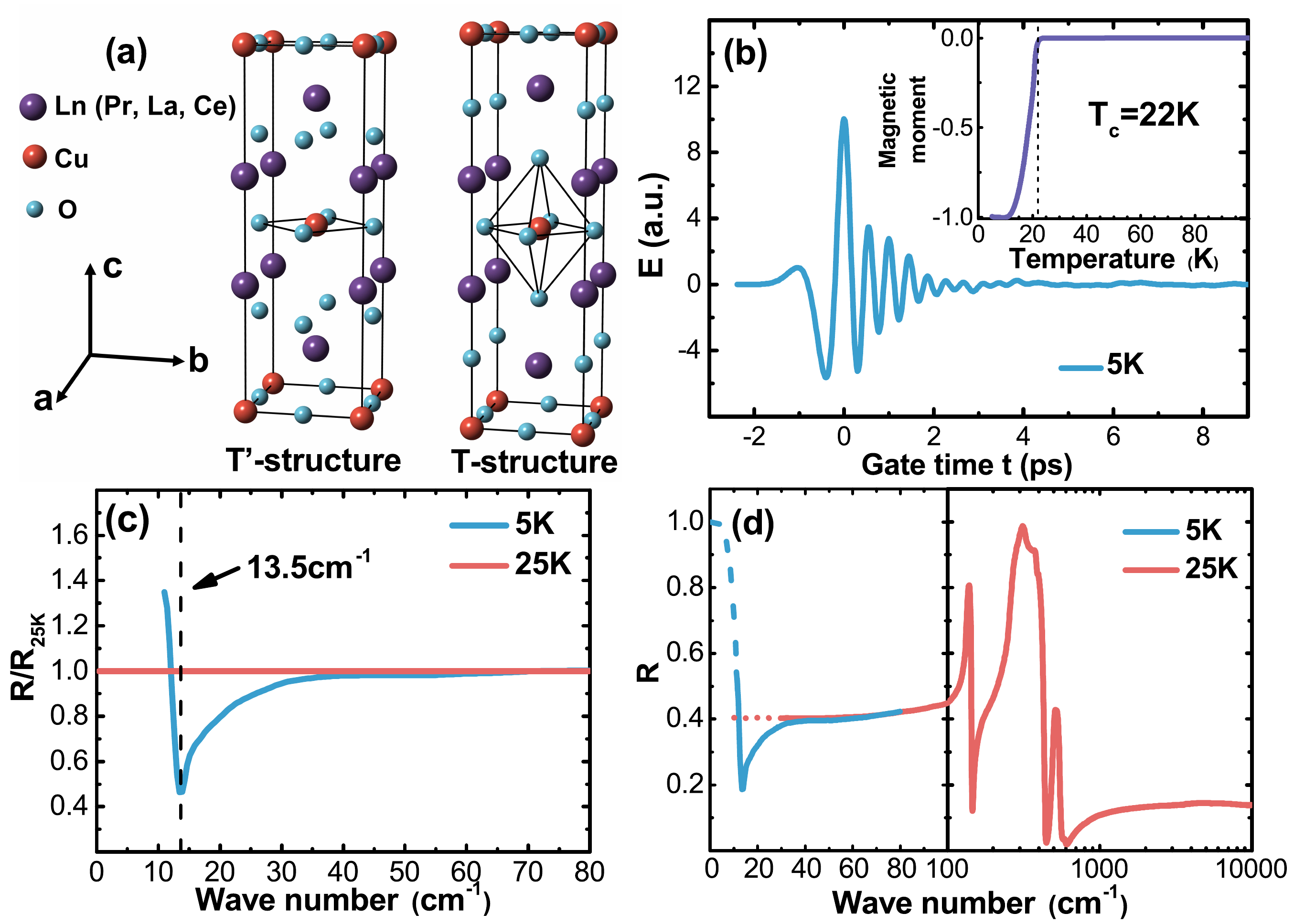}
  \caption{Sample structure and characterization. (a) A comparison between T'- and T-structures. The T'-structure is characterized by a lack of apical oxygen compared with T-structured hole-doped HTSC. (b) Inset: a sharp superconducting transition at T$_c$ = 22 K is determined by temperature dependent magnetic susceptibility measurement. Main panel: the reflected THz electric field of PLCCO along c-axis at 5 K is characterized by a THz time-domain spectrometer. (c) Reflected spectra at 5K (below T$_c$) are normalized to spectra at 25 K (above T$_c$). (Detailed procedure can be found in Method section). A very sharp JPE develops near 13.5\cm at 5 K, which is indicated by a dash line. (d) The broad band reflectivity spectra at 25 K and 5 K are acquired by combining the measurements of FTIR and THz time-domain spectrometers.}\label{Fig:static}
\end{figure*}

Electron-doped High-T$_c$ superconducting cuprates (HTSC), as a family of superconducting copper oxides, is always investigated by comparing with hole-doped one since it was first discovered in 1989 \cite{Tokura1989,PhysRevLett.62.1197}. Electron-doped HTSC can be realized as a system without apical oxygens, which is only in specific ``214'' T'-structure, \emph{ e.g.} Ce-doped Nd$_2$CuO$_4$ \cite{Tokura1989}, or related rare-earth element-based compounds such as Pr$_{2-x}$Ce$_x$CuO$_4$, Pr$_{1-x}$LaCe$_x$CuO$_4$, \textit{etc}. Compared with hole-doped La$_{1-x}$Ba$_x$CuO$_4$ with K$_2$NiF$_4$ structure (T-structure), the out-of-plane oxygen shifts from the apical position to the fixed position (0, 1/2, 1/4). This shift alters the coordinations of both Cu and Ln atoms. Copper becomes strictly square-planar coordinated with no apical oxygens. A comparison between T'- and T-structures is shown in Fig. \ref{Fig:static} (a). Multiple other aspects have already been verified exhibiting both similarities and differences between those two compounds \cite{RN95}.

Ultrafast pump-probe measurements, an emerging technique to measure the non-equilibrium spectroscopy after the excitation of ultrafast pump pulses, have already developed as crucial tools for understanding and manipulating the different degrees of freedom in High-T$_c$ superconducting cuprates (HTSC) \cite{RN180}. Up to now, ultrafast pump-probe measurements performed on electron-doped HTSC were dedicated to disentangle complex degrees of freedom in the system \cite{Liu1993Ultrafast,LONG200659,CAO2008894,RN154,RN183,RN186}, in which the fluence of pump pulses is always quite small and regarded as week external perturbations for material systems. In single-color pump-probe experiments on electron-doped HTSC \cite{Liu1993Ultrafast,LONG200659,CAO2008894,RN154,RN183}, NIR pump pulses with the fluence below 10 $\upmu$J/cm$^{2}$ were expected to suppress/destroy superconductivity by breaking Cooper pairs into quasiparticles. Using some effective phenomenological models such as Rothwarf-Taylor model \cite{PhysRevLett.19.27}, the subsequent decay procedure of those quasiparticles may disclose the complex interplay among electrons, lattice and spin dynamics. In NIR/THz pump-THz probe measurements in ab-plane of Pr$_{1.85}$Ce$_{0.15}$CuO$_{4-\delta}$ \cite{RN186}, the depletion and the recovery of the superconducting state after excited by pump pulses can be observed in the pump-induced change of conductivity.

Distinguished from weak pump pulses expected to excite quasiparticles from occupied state to unoccupied one, intense pump pulses turn to induce a phase transition and drive a material system into a novel phase unaccessible for traditional physical parameters such as temperature and pressure \cite{Stojchevska177, Gedik425, RN211}, which provide unique opportunities to manipulate different orders in material systems. A surprising finding is that intense mid-infrared ultrafast excitation can drive a  non-superconducting hole-doped HTSC into a phase with a Josephson plasmon edge (JPE) in the c-axis optical spectra, which was first observed in a stripe-ordered cuprate at 10 K whose T$_c$ is less than 2 K \cite{Fausti189}, then in underdoped YBa$_2$Cu$_3$O$_{6.5}$ even above room temperature \cite{Kaiser2014,hu2014optically}. Later on, it was found that intense NIR pulses can also induce a phase transition in La$_{1-x}$Ba$_x$CuO$_4$ systems \cite{Nicoletti2014,PhysRevB.91.174502, Zhang2017b}.

Here we report photoexcited c-axis dynamics in an electron-doped cuprate Pr$_{0.88}$LaCe$_{0.12}$CuO$_4$ (PLCCO) with T$_c$=22 K after the excitation of intense NIR pump at 1.28 $\upmu m$ with polarization of $\textbf{E}\parallel c$-axis. We observe a photoexcitation induced split of the JPE in superconducting state. Remarkably, no significant decay is observed up to the longest measurement time delay 210 ps after excitation. The observation indicates clearly that the intense pump with energy much higher than the superconducting pairing energy turns to drive the system from a superconducting state with a uniform Josephson coupling to a new metastable superconducting phase with modulated Josephson coupling strengths, rather than breaking Cooper pairs or excite quasiparticles to unoccupied states far above the Fermi level as weak pump pulses do.

\textbf{Static c-axis THz and infrared reflectivity spectra.}
HTSC are highly anisotropic materials, in which conducting CuO$_2$ layers are usually separated by insulating block layers, leading to insulator-like \emph{dc} and optical responses along c-axis in normal state. The optical reflectivity spectra of PLCCO along c-axis with T$_c$=22 K are acquired by a combination of a time-domain THz spectrometer and Fourier transform infrared spectrometers. Figure \ref{Fig:static} (b) displays the reflected THz electric field $E(t)$ of PLCCO at 5 K in time-domain recorded by electric-optic sampling. The time zero $E$($t$ = 0ps) is denoted as the peak position of $E(t)$, $\textit{i.e.}$, $E_{peak}$. Figure \ref{Fig:static} (c) shows the normalized  reflected spectra below and above T$_c$. The broadband c-axis optical reflectivity spectra are presented in Fig. \ref{Fig:static} (d). The low frequency c-axis optical spectra are dominated by the infrared active phonons with little contribution from free carriers. In the normal state, reflectivity below 80 cm$^{-1}$ (or 2.5 THz) is low and almost featureless, indicating an insulating response. However, when PLCCO goes into superconducting state, a JPE develops in the c-axis reflectivity spectra and lies near 13.5\cm at 5 K, which is caused by the Josephson tunneling effect of condensed superfluid carriers. Manifestation of such c-axis plasma edge is taken as an optical evidence for the occurrence of superconductivity. The reflectivity spectra are similar to those measured on Nd$_{1.85}$Ce$_{0.15}$CuO$_4$ \cite{RN90}.

\begin{figure}[htbp]
\setlength{\abovecaptionskip}{-0.01cm}
\setlength{\belowcaptionskip}{-0.45cm}
  \centering
%Requires \usepackage{graphicx}
\includegraphics[width=12cm]{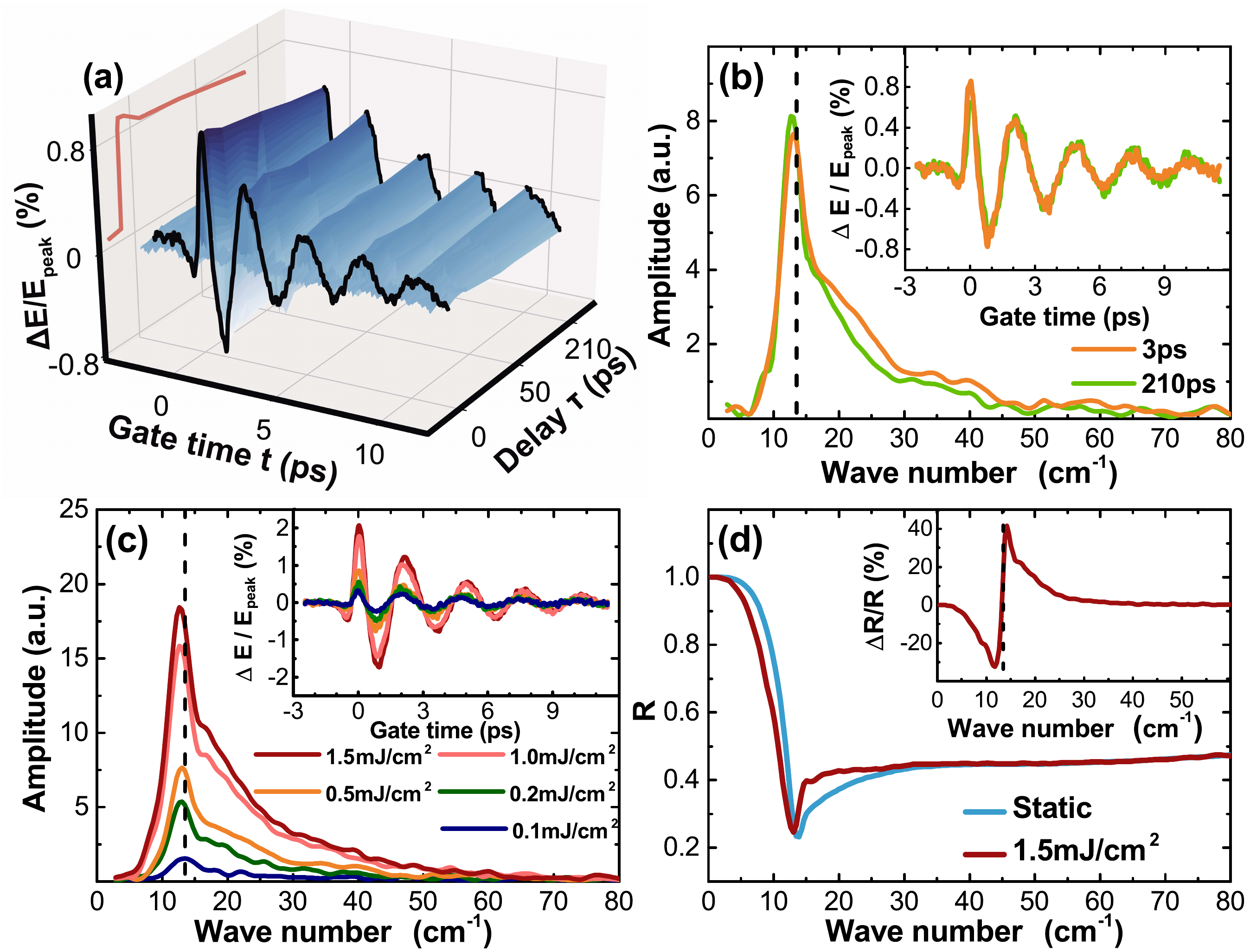}\\
  \caption{Photoexcitation induced changes at 5K. (a) The relative change of  the reflected THz electric field $\Delta E(t, \tau)/E_{peak}$ at 5 K at different decay time $\tau$ after excited by 1.28 $\upmu$m pulses at fluence of 0.5 mJ/cm$^2$. The red line shows the decay of $\Delta E($t$ = 0ps, \tau)/E_{peak}$. The black lines indicate $\Delta E(t, \tau = 3$ps$)/E_{peak}$ and $\Delta E(t, \tau = 210$ps$)/E_{peak}$. (b) Inset: the photoexcitation induced relative change $\Delta E(t, \tau)/E_{peak}$ in time domain at $\tau$=3 ps and 210 ps. Main panel: the Fourier transformed spectrum of $\Delta E(\omega, \tau)$. (c) Inset: fluence dependence of  $\Delta E(t, \tau)/E_{peak}$. Main panel: the Fourier transformed spectrum. (d) R($\omega)$ before and after excitation by 1.5 mJ/cm$^2$ pulses. Inset shows the ratio of the reflectivity change relative to the static values. The penetration depth mismatch is not considered here.}\label{Fig:DeltaE}
\end{figure}

\textbf{Photoexcitation induced changes in THz spectra.}
Figure \ref{Fig:DeltaE} (a) illustrates the relative changes of  the reflected THz electric field $\Delta E(t, \tau)/E_{peak}$ at 5 K at different decay time $\tau$ after excited by 1.28 $\upmu$m pulses at fluence of 0.5 mJ/cm$^2$. We define time zero of the decay procedure $\tau$=0 ps at the position where $\Delta E(t$ = 0 ps, $\tau)/E_{peak}$ starts to change, so the state before $\tau$=0 ps is defined as ``static'' state. The red line shows the decay of $\Delta E($t$ = 0ps, \tau)/E_{peak}$. The black lines indicate the relative change of THz electric field at $\tau$=3 ps and 210 ps respectively, $\textit{i.e.}$ $\Delta E(t, \tau = 3 $ ps$)/E_{peak}$ and $\Delta E(t, \tau = 210 $ ps$)/E_{peak}$, which is also presented in the inset of Fig. \ref{Fig:DeltaE} (b) more clearly. Roughly a 0.8$\%$ maximum relative change is seen within 3 ps after excitation and clear oscillations can be seen in $\Delta E(t, \tau)/E_{peak}$, which gives a peak slightly below 13.5 \cm in frequency domain $\Delta E(\omega,\tau)$ after Fourier transformation, as shown in the main panel of Fig. \ref{Fig:DeltaE} (b). The signal does not show significant decay up to the longest measured time delay 210 ps. This pronounced peak in $\Delta E(\omega, \tau)$ suggests that photoexcitation induced change occurs predominantly near the static JPE position. We also performed NIR pump THz probe measurements at different pump fluence as shown in Fig. \ref{Fig:DeltaE} (c). It can be seen that the pump induced change gets more and more significant as increasing pump fluence. Figure \ref{Fig:DeltaE} (d)  shows the reflectivity spectrum calculated from complex reflection coefficient \cite{Zhang2017b} at $\tau$=3 ps after excited by 1.5 mJ/cm$^2$ pulses. The static reflectivity is also plotted for comparison. The main changes after excitation are the slight redshift of static JPE and the lifting of reflectivity above static JPE position. Reflectivity spectra at $\tau$=3 ps merges into the static one above 30 \cm.  Those features can be more clearly observed in the ratio of reflectivity change over the static values, as shown in the inset of Fig. \ref{Fig:DeltaE} (d).

\begin{figure}[htbp]
%%\begin{tabular}{cc}
\begin{minipage}{0.33\linewidth}
\centerline{\includegraphics[width=6cm]{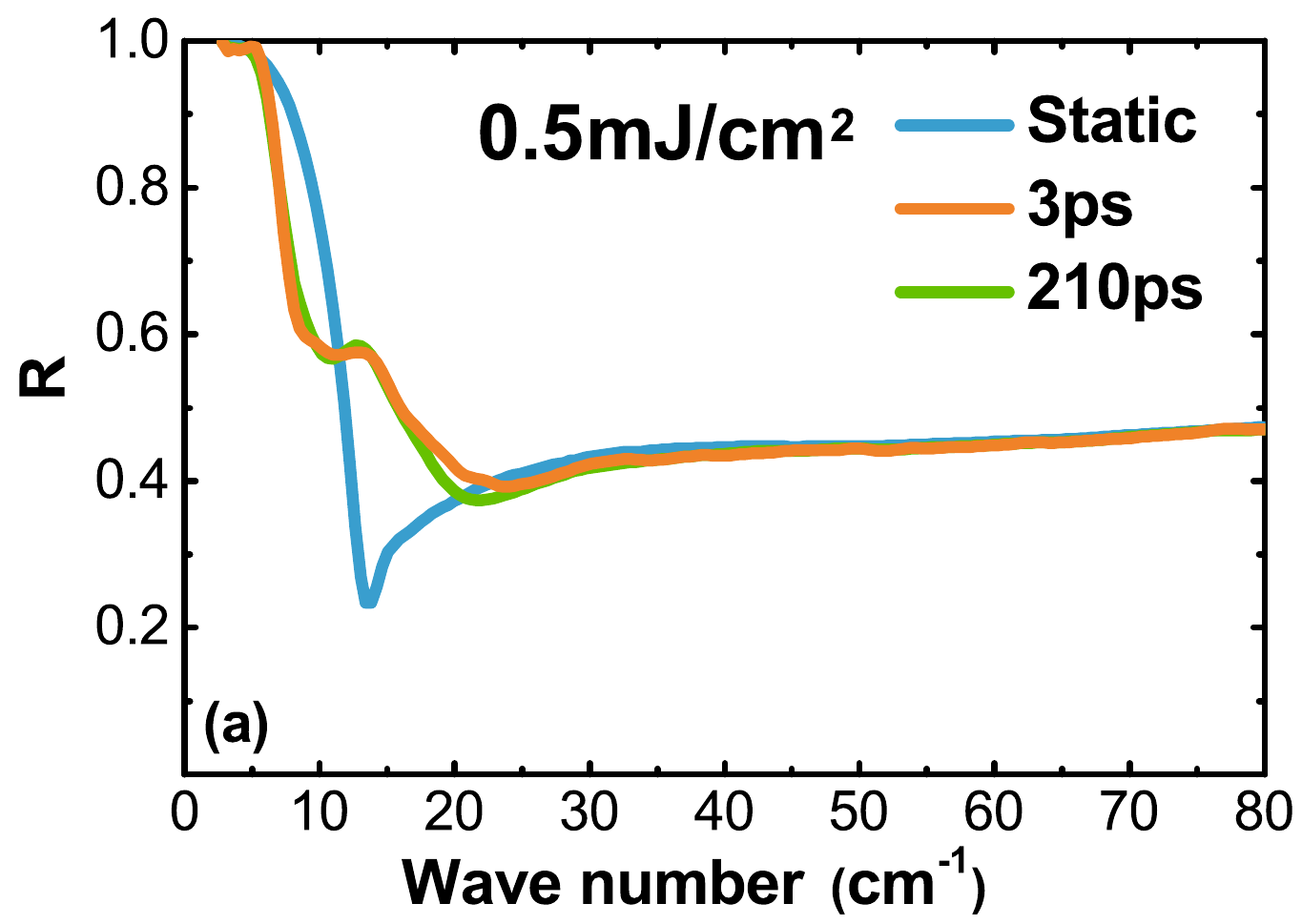}}
\end{minipage}
\hfill
\begin{minipage}{.33\linewidth}
\centerline{\includegraphics[width=6cm]{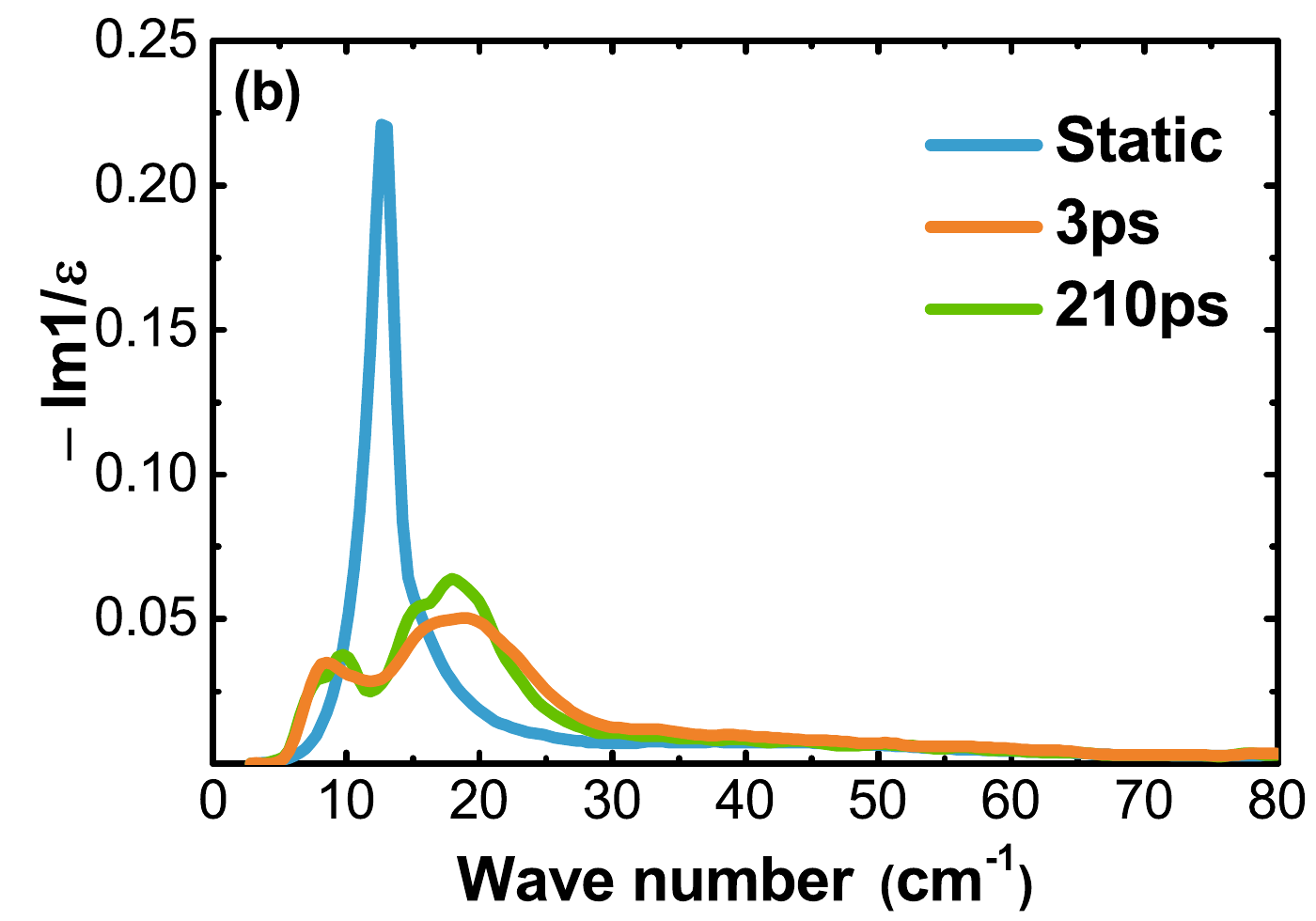}}
\end{minipage}
\hfill
\begin{minipage}{.33\linewidth}
\centerline{\includegraphics[width=6cm]{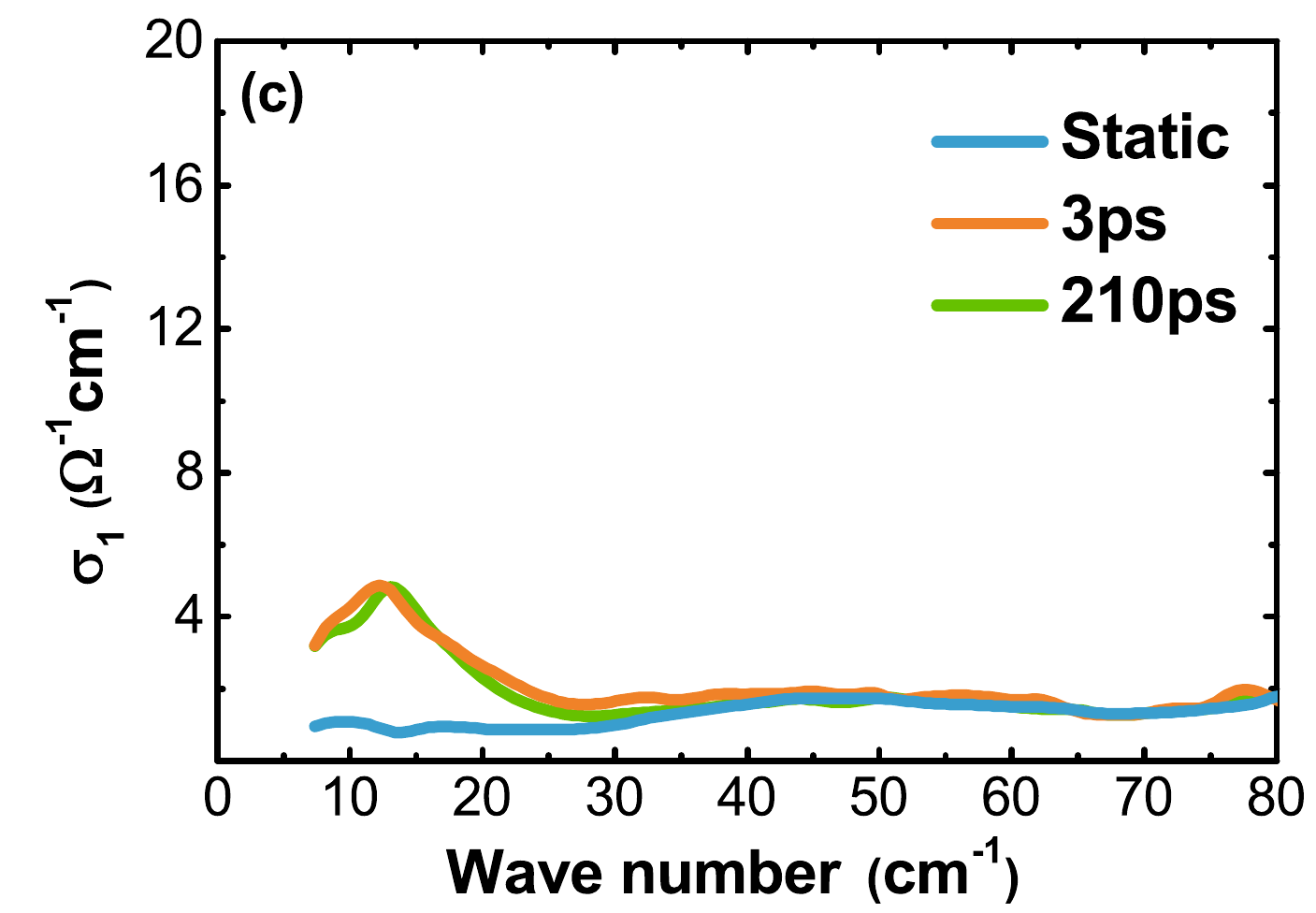}}
\end{minipage}
\vfill
\begin{minipage}{0.33\linewidth}
\centerline{\includegraphics[width=6.0cm]{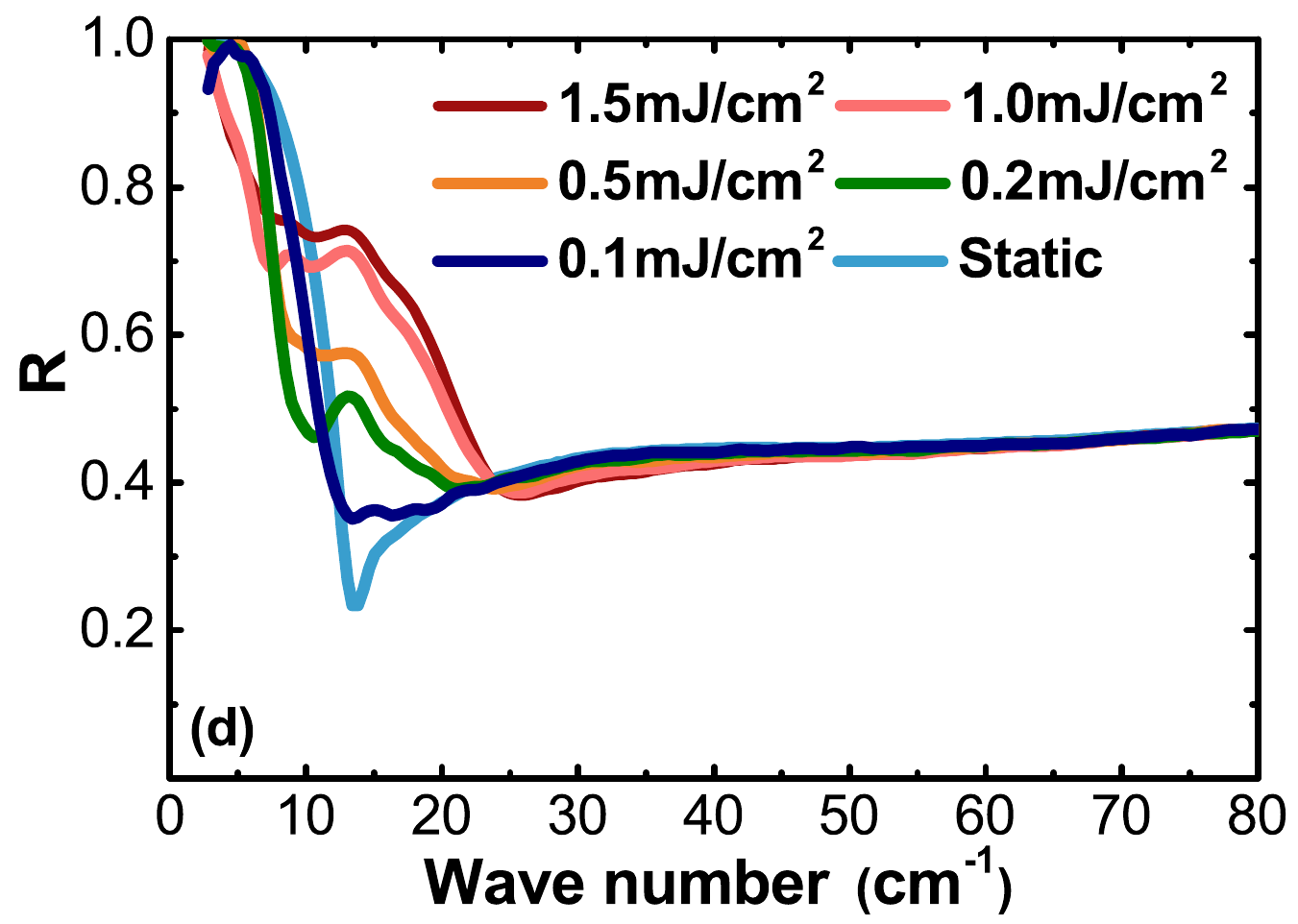}}
\end{minipage}
\hfill
\begin{minipage}{0.33\linewidth}
\centerline{\includegraphics[width=6.0cm]{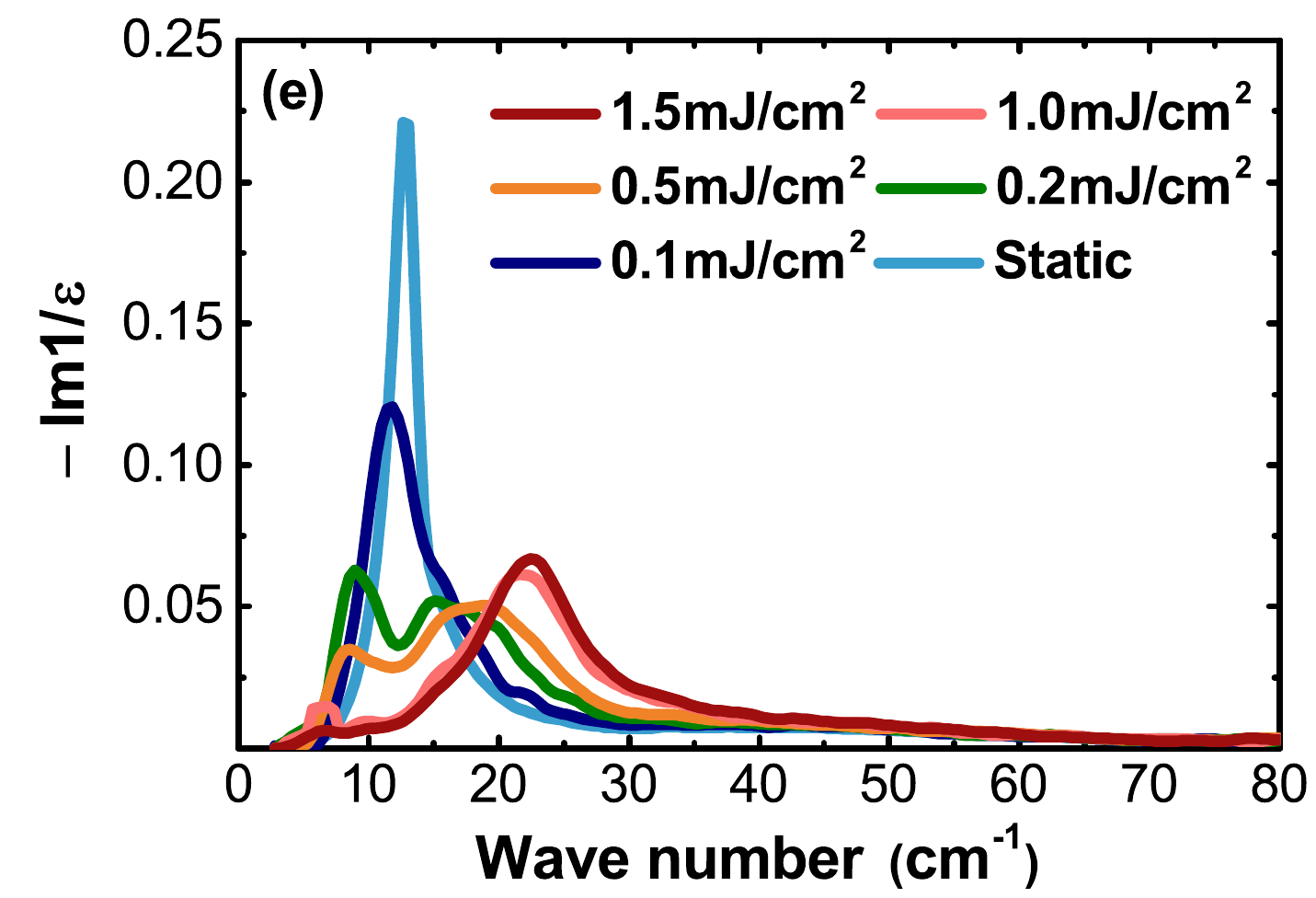}}
\end{minipage}
\hfill
\begin{minipage}{0.33\linewidth}
\centerline{\includegraphics[width=6.0cm]{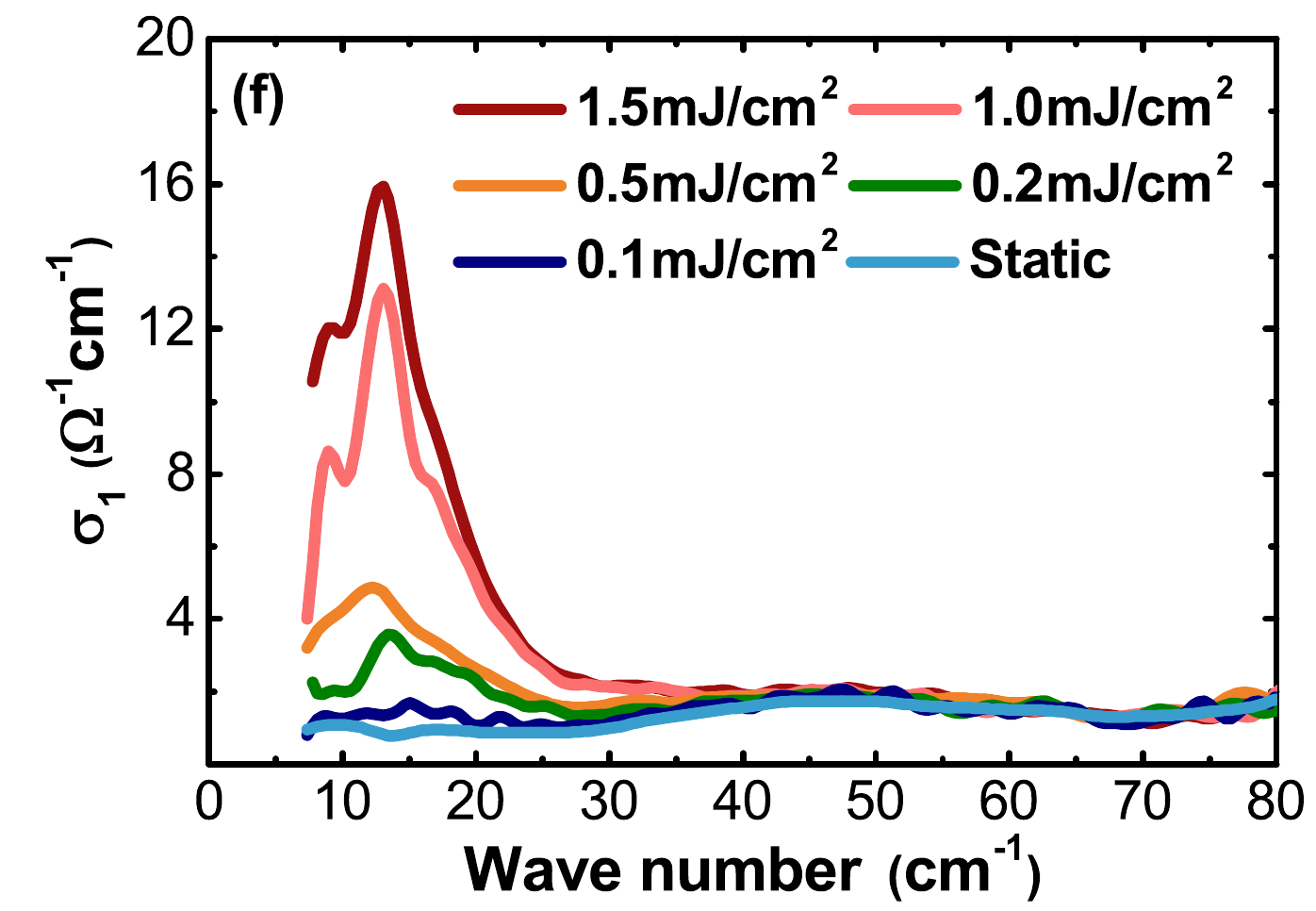}}
\end{minipage}
%\end{tabular}
\caption{Calculated optical constants from a multilayer model. (a)-(c) show the temporal evolution of calculated R($\omega$), energy loss function Im(-1/$\varepsilon(\omega$)) and real part of conductivity spectra $\sigma_1(\omega)$ after photo-excitations. No significant change can be found in the decay procedure. (d)-(f) show the fluence-dependence of photoexcitation induced values. The photoexitation induced change gets more and more significant as increasing pump fluence. }
\label{Fig:mismatch}
\end{figure}

In Fig. \ref{Fig:mismatch}, a multilayer model is used for calculating the photoexcitation induced change of  reflectivity R($\omega$), energy loss function Im(-1/$\varepsilon(\omega$)) and real part of conductivity spectra $\sigma_1(\omega)$, to eliminate the effects of penetration depths mismatch between 1.28$\upmu$m pump pulses and THz probe pulses (wavelength range from 120 $\upmu$m to 1000 $\upmu$m) \cite{Zhang2017b}. We first examine R($\omega$) at two representative time delays after excitation. At the maximum photoexcitation induced signal position, \emph{i.e.} $\tau$=3 ps (Fig. \ref{Fig:mismatch} (a)), two edges can be seen in R($\omega$), which locate just below ($\sim$ 8 \cm) and above ($\sim$ 20 \cm) the static JPE position. They can be identified more clearly as peaks in Im(-1/$\varepsilon(\omega$)) as presented in Fig. \ref{Fig:mismatch} (b).  Compared with the static state with only one longitudinal Josephson plasmon mode, the photoexcitation induced effect is the splitting of the peak in Im(-1/$\varepsilon(\omega$)), which represents the emergence of two longitudinal Josephson plasmon modes with modulated Josephson coupling strengths. Presence of two longitudinal Josephson plasmon modes would result in the formation of a transverse Josephson plasmon mode, which can be seen as a peak in $\sigma_1(\omega)$ as shown in Fig. \ref{Fig:mismatch} (c). This transverse Josephson plasmon mode can be regarded as an out-of-phase oscillation of the two individual  longitudinal Josephson plasmon modes \cite{VanderMarel1996} and has been observed in many cuprate systems\cite{Shibata1998,Grueninger1999,Timusk2003,Tajima2012}. Similar effects can be also observed at $\tau$=210 ps and there are no distinct difference between these two time delays, which indicates that PLCCO may have been driven into a metastable state by the pump pulses.

Figure \ref{Fig:mismatch} (d)-(f) show the fluence dependence of photoexcitation induced change at $\tau$=3 ps. When the pump fluence is quite small, \emph{e.g.} 0.1 mJ/cm$^{2}$, only a slight redshift of static JPE and a lifting of reflectivity above static JPE position can be observed in R($\omega$). With increasing the pump fluence,  the static JPE are suppressed to lower energy scale and a new edge above static JPE position with increasingly higher energy scale can be observed as shown in Fig. \ref{Fig:mismatch} (d). Figure \ref{Fig:mismatch} (e) shows the corresponding energy loss function. When the pump fluence are tuned from 0.1 mJ/cm$^{2}$ to 0.5 mJ/cm$^{2}$, the separation of the two peaks in Im(-1/$\varepsilon(\omega$))  get more and more unambiguous and significant. Meanwhile, the photoexcitation induced new longitudinal mode at higher energy gets more pronounced as increasing the pump fluence. When fluence goes above 1 mJ/cm$^{2}$, the edge at lower frequency in reflectance R($\omega$) becomes less pronounced, but the edge at higher frequency become more eminent. Then the peak feature in Im(-1/$\varepsilon(\omega$)) corresponding to lower edge becomes almost invisible, which may result from the measurement limitation of THz pulses generated by ZnTe crysals. Meanwhile, the peak in $\sigma_1(\omega)$ gets more and more significant as shown in Fig. \ref{Fig:mismatch} (f). The most prominent result of our experiments is the observation of photo-induced two long-lived longitudinal Josephson plasmon modes and a transverse plasmon mode. It suggests the development of two inequivalent Josephson couplings along the c-axis.

\textbf{ Discussions and implications.}
The experimental result is highly nontrivial for two reasons. Firstly, the photon energy of NIR pump is much higher than the superconducting energy gap 2$\Delta$, it is expected that the intense pump would break Cooper pairs and destroy superconductivity. However, this is not the case. The compound is still superconducting, nevertheless with modulated Josephson coupling strengths. Secondly, the NIR pump should excite quasiparticles from occupied state to unoccupied state far above the Fermi level, so after excitations one would normally expect to detect the nonequilibrium dynamics of those excited quasiparticles relaxing towards equilibrium state via different quasiparticle-bosonic excitation interactions. This appears also not the case, or at most a minor effect. The true and dominant effect of intense pump is to drive the system to a new metastable state that does not exist before excitations.

\begin{figure}[htbp]
\setlength{\abovecaptionskip}{-0.005cm}
\setlength{\belowcaptionskip}{-0.4cm}
  \centering
  % Requires \usepackage{graphicx}
  \includegraphics[width=12cm]{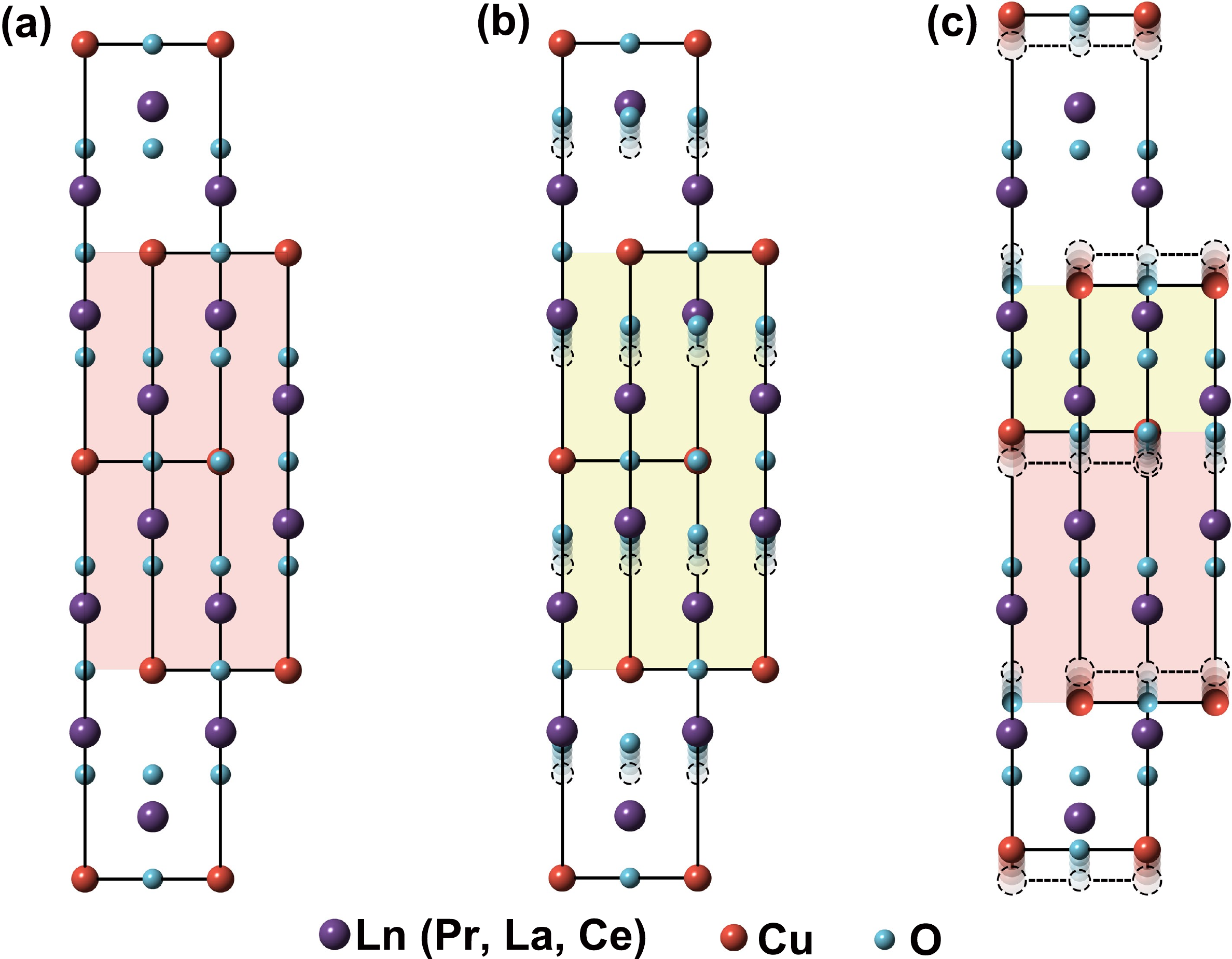}\\
  \caption{A possible scenario for the presence of inequivalent coupling strength between different CuO$_2$ planes. (a) Electron-doped HTSC is characterized by a lack of apical oxygens, which are always related to photoexcitation induced phase transition in hole-doped HTSC. (b) Inequivalent coupling strengths can not be obtained by the deviation of oxygen at (0, 1/2, 1/4) position from its equilibrium position. (c) Two different Josephson coupling strengths (shaded in different colors) can be induced by speculating the displacement of two sub-tetragonal lattices.}\label{Fig:PumpStruc}
\end{figure}

Josephson coupling with multiple coupling strengths primarily comes from either the crystal structure with various kinds of Josephson junctions \cite{Shibata1998,Grueninger1999,Timusk2003,Tajima2012} or the external magnetic field induced effect \cite{PhysRevLett.89.247001,PhysRevB.76.054524}. Nevertheless, it seems impossible for the magnetic-field component of pump pulses with 35 fs pulse duration to induce long-lived Josephson vortices in alternate insulating layers lasting for no less than 210 ps, which is the origin of external magnetic field induced modulated Josephson coupling strengths \cite{PhysRevLett.89.247001,PhysRevB.76.054524}. Since the system goes to a new state in which the physical properties do not change appreciably with time delay, we have to consider subtle structural phase transition induced by the intense pump. In fact, photoinduced structural phase transitions have been found in different compounds \cite{Rini2007b,VanceR2014,Gedik2007,Ichikawa2011}. For hole-doped La$_{1.905}$Ba$_{0.095}$CuO$_4$ superconductor, it is suggested that the intense pump pulse could drive the out-of-plane apical oxygens to deviate from their equilibrium positions, resulting in a modulation of Cu-apical oxygen bond lengths between different CuO$_2$ planes, and therefore two different Josephson coupling strengths \cite{Zhang2017b}. For the electron-doped cuprate, the Cu is strictly square-planar coordinated with no apical oxygen, as shown in Fig. \ref{Fig:PumpStruc} (a). Even if we assume that the pump can drive the oxygen at (0, 1/2, 1/4) position to deviate from its equilibrium position either downward or upward, it still can not result in inequivalent coupling strengths between neighbouring two CuO$_2$ planes in T'-structure as displayed in Fig. \ref{Fig:PumpStruc} (b). This is very different from the T-structure of hole-doped cuprates with the presence of apical oxygens. To obtain a modulation of different coupling strengths, we have to assume that the CuO$_2$ planes would be affected by the intense pump \cite{RN138}. Like the T-structure of hole-doped cuprate, the T'-structure also has a body centered tetragonal structure which can be considered as two sub-tetragonal lattices shifting relatively with a wave vector of (1/2,1/2,1/2) in real space. If we assume that the intense pump can cause two sub-tetragonal lattices to have a small displacement along c-axis relative to the static state, then the spacing modulation between neighboring CuO$_2$ planes could develop, as illustrated in Fig. \ref{Fig:PumpStruc} (c). As a result, two different Josephson coupling strengths could be induced. This would explain the formation of two longitudinal Josephson plasmons and a transverse mode. At present, this is purely a speculation. Apparently, further detailed studies on the photoinduced structural change and its decay dynamics by other time resolved probes below T$_c$ are needed to verify this scenario.

The measurement results may shed new light in understanding the photoinduced phase transition phenomena in cuprates and other strongly correlated materials. A leading interpretation for the photoinduced  phase transition phenomena in cuprates and related materials is so-called phonon pumping \cite{Fausti189,Kaiser2014,hu2014optically}. Key to those experiments is the resonant pumping of relevant infrared active CuO mode being associated with either the apical oxygen or oxygen within CuO$_2$ plane. Since the pumping energy about 1 eV in the present study is much higher than the highest CuO phonon mode (which is lower than 0.1 eV), resonant phonon pumping is unambiguously ruled out. Nevertheless, the significant photoinduced effect revealed in this work indicates that the NIR pumping can also lead to distortion or displacement of the lattice structure through nonlinear phononics.

To summarize, we performed NIR pump c-axis THz probe measurement on a superconducting single crystal PLCCO with T$_c$=22 K. The intense pump induces a splitting of Josephson plasma edge below T$_c$. The photoexcitation induced spectral change does not exhibit observable decay up to the longest measured time delay 210 ps. As increasing the pump fluence, the splitting effect gets more significant. The measurement reveals that intense NIR pump drives the system from an equilibrium superconducting state with uniform Josephson coupling strength to a new metastable superconducting phase with modulated Josephson coupling strengths, rather than destroying superconductivity or exciting quasiparticles to unoccupied states far above the Fermi level. We speculate that the intense near infrared pump induces certain sort of displacement of the lattice structure.

\textbf{Methods}

\begin{small}

\textbf{Sample preparation and characterization}. Single crystals of PLCCO were grown using the travelling-solvent floating-zone technique. A post annealing process is conducted to remove excess oxygen and induce superconductivity\cite{Li2008}. The temperature dependent magnetic susceptibility is measured by Quantum Design physical property measurement system.

\textbf{Equilibrium optical spectra}. The broadband c-axis optical reflectivity spectra were acquired by Fourier transform infrared spectrometers (FTIR) (Bruker 113v and Vertex 80v, ranging from 30 \cm to15000 \cm) using a in-situ gold overcoating technique. Limited by the signal-to-noise ratio in low frequency range of FTIR, a THz time-domain spectrometer (ranging from 10 \cm to 80 \cm) is used to determine the ratios of reflectivity below and above T$_c$, \textit{i.e. }R$_{5K}(\omega)$/R$_{25K}(\omega)$. According to the definition of reflectivity R($\omega$)=$|\tilde{r}(\omega)|^{2}$=$|\tilde{E}_{reflected}(\omega)/\tilde{E}_{incident}(\omega)|^{2}$, the relative refelctivity ratio of different temperature R$_{5K}(\omega)$/R$_{25K}(\omega)$=$|\tilde{E}_{5K}(\omega)/\tilde{E}_{25K}(\omega)|^{2}$ can be acquired by measuring the strengths of  reflected THz electric field $\tilde{E}_{5K}(\omega)$ and $\tilde{E}_{25K}(\omega)$ using THz time-domain spectrometer. The low frequency below 30 \cm of  R$_{25K}(\omega)$ is extrapolated in a constant form. R$_{5K}(\omega)$ can be acquired according to  the ratios R$_{5K}(\omega)$/R$_{25K}(\omega)$. The low frequency below 10 \cm is extrapolated using R$(\omega)\varpropto1-\omega^{4}$.

\textbf{Photoexcitation induced changes in THz spectral.} The photoexcitation induced change of c-axis reflectivity was measured by a tunable NIR to MIR pump-THz probe spectroscopy system, which is constructed based on an amplified Ti:sapphire laser system producing 800 nm, 35 fs pulses at a 1 kHz repetition rate and a two-output optical parametric amplifier. Detailed experimental setup and measurement techniques are presented elsewhere\cite{Zhang2017,Zhang2017b}.

\textbf{Multilayer model used for transient optical constants calculation}. The nonnegligible penetration depth mismatch between NIR pump (at 1.28 $\upmu$m) and the THz probe pulses (below 2.5 THz) should be taken into account to derive the true pump-induced THz spectral change. We use a multilayer model with the incident angle of 30$\degree$ to calculate the pump-induced transient optical constants. The detailed procedure can be found in the supplementary imformation of another article\cite{Zhang2017b}.

\end{small}

%\bibliographystyle{nc}
%\bibliography{Terahertz}

\begin{thebibliography}{35}%
\makeatletter
\providecommand \@ifxundefined [1]{%
 \@ifx{#1\undefined}
}%
\providecommand \@ifnum [1]{%
 \ifnum #1\expandafter \@firstoftwo
 \else \expandafter \@secondoftwo
 \fi
}%
\providecommand \@ifx [1]{%
 \ifx #1\expandafter \@firstoftwo
 \else \expandafter \@secondoftwo
 \fi
}%
\providecommand \natexlab [1]{#1}%
\providecommand \enquote  [1]{``#1''}%
\providecommand \bibnamefont  [1]{#1}%
\providecommand \bibfnamefont [1]{#1}%
\providecommand \citenamefont [1]{#1}%
\providecommand \href@noop [0]{\@secondoftwo}%
\providecommand \href [0]{\begingroup \@sanitize@url \@href}%
\providecommand \@href[1]{\@@startlink{#1}\@@href}%
\providecommand \@@href[1]{\endgroup#1\@@endlink}%
\providecommand \@sanitize@url [0]{\catcode `\\12\catcode `\$12\catcode
  `\&12\catcode `\#12\catcode `\^12\catcode `\_12\catcode `\%12\relax}%
\providecommand \@@startlink[1]{}%
\providecommand \@@endlink[0]{}%
\providecommand \url  [0]{\begingroup\@sanitize@url \@url }%
\providecommand \@url [1]{\endgroup\@href {#1}{\urlprefix }}%
\providecommand \urlprefix  [0]{URL }%
\providecommand \Eprint [0]{\href }%
\providecommand \doibase [0]{http://dx.doi.org/}%
\providecommand \selectlanguage [0]{\@gobble}%
\providecommand \bibinfo  [0]{\@secondoftwo}%
\providecommand \bibfield  [0]{\@secondoftwo}%
\providecommand \translation [1]{[#1]}%
\providecommand \BibitemOpen [0]{}%
\providecommand \bibitemStop [0]{}%
\providecommand \bibitemNoStop [0]{.\EOS\space}%
\providecommand \EOS [0]{\spacefactor3000\relax}%
\providecommand \BibitemShut  [1]{\csname bibitem#1\endcsname}%
\let\auto@bib@innerbib\@empty
%</preamble>
\bibitem [{\citenamefont {Tokura}, \citenamefont {Takagi},\ and\ \citenamefont
  {Uchida}(1989)}]{Tokura1989}%
  \BibitemOpen
  \bibinfo {author} {Tokura Y}, \bibinfo {author} {Takagi H}\ and\ \bibinfo
  {author} {Uchida S},\ \href {\doibase 10.1038/337345a0} {\enquote {\bibinfo
  {title} {{A superconducting copper oxide compound with electrons as the
  charge carriers}},}\ } (\bibinfo {year} {1989})\BibitemShut {NoStop}%
\bibitem [{\citenamefont {Takagi}, \citenamefont {Uchida},\ and\ \citenamefont
  {Tokura}(1989)}]{PhysRevLett.62.1197}%
  \BibitemOpen
  \bibinfo {author} {Takagi H}, \bibinfo {author} {Uchida S}\ and\ \bibinfo
  {author} {Tokura Y},\ \href {\doibase 10.1103/PhysRevLett.62.1197} {\bibfield
   {journal} {\bibinfo  {journal} {Phys. Rev. Lett.}\ }\textbf {\bibinfo
  {volume} {62}},\ \bibinfo {pages} {1197} (\bibinfo {year}
  {1989})}\BibitemShut {NoStop}%
\bibitem [{\citenamefont {Armitage}, \citenamefont {Fournier},\ and\
  \citenamefont {Greene}(2010)}]{RN95}%
  \BibitemOpen
  \bibinfo {author} {Armitage N~P}, \bibinfo {author} {Fournier P}\ and\
  \bibinfo {author} {Greene R~L},\ \href {\doibase 10.1103/RevModPhys.82.2421}
  {\bibfield  {journal} {\bibinfo  {journal} {Reviews of Modern Physics}\
  }\textbf {\bibinfo {volume} {82}},\ \bibinfo {pages} {2421} (\bibinfo {year}
  {2010})}\BibitemShut {NoStop}%
\bibitem [{\citenamefont {Giannetti}\ \emph {et~al.}(2016)\citenamefont
  {Giannetti}, \citenamefont {Capone}, \citenamefont {Fausti}, \citenamefont
  {Fabrizio}, \citenamefont {Parmigiani},\ and\ \citenamefont
  {Mihailovic}}]{RN180}%
  \BibitemOpen
  \bibinfo {author} {Giannetti C}, \bibinfo {author} {Capone M}, \bibinfo
  {author} {Fausti D}, \bibinfo {author} {Fabrizio M}, \bibinfo {author}
  {Parmigiani F}\ and\ \bibinfo {author} {Mihailovic D},\ \href {\doibase
  10.1080/00018732.2016.1194044} {\bibfield  {journal} {\bibinfo  {journal}
  {Advances in Physics}\ }\textbf {\bibinfo {volume} {65}},\ \bibinfo {pages}
  {58} (\bibinfo {year} {2016})}\BibitemShut {NoStop}%
\bibitem [{\citenamefont {Liu}\ \emph {et~al.}(1993)\citenamefont {Liu},
  \citenamefont {Whitaker}, \citenamefont {Uher}, \citenamefont {Peng},
  \citenamefont {Li},\ and\ \citenamefont {Greene}}]{Liu1993Ultrafast}%
  \BibitemOpen
  \bibinfo {author} {Liu Y}, \bibinfo {author} {Whitaker J~F}, \bibinfo
  {author} {Uher C}, \bibinfo {author} {Peng J}, \bibinfo {author} {Li Z~Y}\
  and\ \bibinfo {author} {Greene R~L},\ \href@noop {} {\bibfield  {journal}
  {\bibinfo  {journal} {Applied Physics Letters}\ }\textbf {\bibinfo {volume}
  {63}},\ \bibinfo {pages} {979} (\bibinfo {year} {1993})}\BibitemShut
  {NoStop}%
\bibitem [{\citenamefont {Long}\ \emph {et~al.}(2006)\citenamefont {Long},
  \citenamefont {Zhao}, \citenamefont {Zhao}, \citenamefont {Qiu},
  \citenamefont {Zhang}, \citenamefont {Fu}, \citenamefont {Wang},
  \citenamefont {Zhang}, \citenamefont {Zhao}, \citenamefont {Yang},\ and\
  \citenamefont {Wang}}]{LONG200659}%
  \BibitemOpen
  \bibinfo {author} {Long Y}, \bibinfo {author} {Zhao L}, \bibinfo {author}
  {Zhao B}, \bibinfo {author} {Qiu X}, \bibinfo {author} {Zhang C}, \bibinfo
  {author} {Fu P}, \bibinfo {author} {Wang L}, \bibinfo {author} {Zhang Z},
  \bibinfo {author} {Zhao S}, \bibinfo {author} {Yang Q}\ and\ \bibinfo
  {author} {Wang G},\ \href {\doibase
  https://doi.org/10.1016/j.physc.2005.12.056} {\bibfield  {journal} {\bibinfo
  {journal} {Physica C: Superconductivity}\ }\textbf {\bibinfo {volume}
  {436}},\ \bibinfo {pages} {59 } (\bibinfo {year} {2006})}\BibitemShut
  {NoStop}%
\bibitem [{\citenamefont {Cao}\ \emph {et~al.}(2008)\citenamefont {Cao},
  \citenamefont {Long}, \citenamefont {Zhang}, \citenamefont {Yuan},
  \citenamefont {Gao}, \citenamefont {Zhao}, \citenamefont {Zhao},
  \citenamefont {Yang}, \citenamefont {Zhao},\ and\ \citenamefont
  {Fu}}]{CAO2008894}%
  \BibitemOpen
  \bibinfo {author} {Cao N}, \bibinfo {author} {Long Y}, \bibinfo {author}
  {Zhang Z}, \bibinfo {author} {Yuan J}, \bibinfo {author} {Gao L}, \bibinfo
  {author} {Zhao B}, \bibinfo {author} {Zhao S}, \bibinfo {author} {Yang Q},
  \bibinfo {author} {Zhao J}\ and\ \bibinfo {author} {Fu P},\ \href {\doibase
  https://doi.org/10.1016/j.physc.2008.02.004} {\bibfield  {journal} {\bibinfo
  {journal} {Physica C: Superconductivity}\ }\textbf {\bibinfo {volume}
  {468}},\ \bibinfo {pages} {894 } (\bibinfo {year} {2008})}\BibitemShut
  {NoStop}%
\bibitem [{\citenamefont {Hinton}\ \emph {et~al.}(2013)\citenamefont {Hinton},
  \citenamefont {Koralek}, \citenamefont {Yu}, \citenamefont {Motoyama},
  \citenamefont {Lu}, \citenamefont {Vishwanath}, \citenamefont {Greven},\ and\
  \citenamefont {Orenstein}}]{RN154}%
  \BibitemOpen
  \bibinfo {author} {Hinton J~P}, \bibinfo {author} {Koralek J~D}, \bibinfo
  {author} {Yu G}, \bibinfo {author} {Motoyama E~M}, \bibinfo {author} {Lu
  Y~M}, \bibinfo {author} {Vishwanath A}, \bibinfo {author} {Greven M}\ and\
  \bibinfo {author} {Orenstein J},\ \href {\doibase
  10.1103/PhysRevLett.110.217002} {\bibfield  {journal} {\bibinfo  {journal}
  {Phys Rev Lett}\ }\textbf {\bibinfo {volume} {110}},\ \bibinfo {pages}
  {217002} (\bibinfo {year} {2013})}\BibitemShut {NoStop}%
\bibitem [{\citenamefont {Vishik}\ \emph {et~al.}(2017)\citenamefont {Vishik},
  \citenamefont {Mahmood}, \citenamefont {Alpichshev}, \citenamefont {Gedik},
  \citenamefont {Higgins},\ and\ \citenamefont {Greene}}]{RN183}%
  \BibitemOpen
  \bibinfo {author} {Vishik I~M}, \bibinfo {author} {Mahmood F}, \bibinfo
  {author} {Alpichshev Z}, \bibinfo {author} {Gedik N}, \bibinfo {author}
  {Higgins J}\ and\ \bibinfo {author} {Greene R~L},\ \href {\doibase
  10.1103/PhysRevB.95.115125} {\bibfield  {journal} {\bibinfo  {journal}
  {Physical Review B}\ }\textbf {\bibinfo {volume} {95}} (\bibinfo {year}
  {2017}),\ 10.1103/PhysRevB.95.115125}\BibitemShut {NoStop}%
\bibitem [{\citenamefont {Beck}\ \emph {et~al.}(2017)\citenamefont {Beck},
  \citenamefont {Klammer}, \citenamefont {Rousseau}, \citenamefont {Obergfell},
  \citenamefont {Leiderer}, \citenamefont {Helm}, \citenamefont {Kabanov},
  \citenamefont {Diamant}, \citenamefont {Rabinowicz}, \citenamefont {Dagan},\
  and\ \citenamefont {Demsar}}]{RN186}%
  \BibitemOpen
  \bibinfo {author} {Beck M}, \bibinfo {author} {Klammer M}, \bibinfo {author}
  {Rousseau I}, \bibinfo {author} {Obergfell M}, \bibinfo {author} {Leiderer
  P}, \bibinfo {author} {Helm M}, \bibinfo {author} {Kabanov V~V}, \bibinfo
  {author} {Diamant I}, \bibinfo {author} {Rabinowicz A}, \bibinfo {author}
  {Dagan Y}\ and\ \bibinfo {author} {Demsar J},\ \href {\doibase
  10.1103/PhysRevB.95.085106} {\bibfield  {journal} {\bibinfo  {journal}
  {Physical Review B}\ }\textbf {\bibinfo {volume} {95}} (\bibinfo {year}
  {2017}),\ 10.1103/PhysRevB.95.085106}\BibitemShut {NoStop}%
\bibitem [{\citenamefont {Rothwarf}\ and\ \citenamefont
  {Taylor}(1967)}]{PhysRevLett.19.27}%
  \BibitemOpen
  \bibinfo {author} {Rothwarf A}\ and\ \bibinfo {author} {Taylor B~N},\ \href
  {\doibase 10.1103/PhysRevLett.19.27} {\bibfield  {journal} {\bibinfo
  {journal} {Phys. Rev. Lett.}\ }\textbf {\bibinfo {volume} {19}},\ \bibinfo
  {pages} {27} (\bibinfo {year} {1967})}\BibitemShut {NoStop}%
\bibitem [{\citenamefont {Stojchevska}\ \emph {et~al.}(2014)\citenamefont
  {Stojchevska}, \citenamefont {Vaskivskyi}, \citenamefont {Mertelj},
  \citenamefont {Kusar}, \citenamefont {Svetin}, \citenamefont {Brazovskii},\
  and\ \citenamefont {Mihailovic}}]{Stojchevska177}%
  \BibitemOpen
  \bibinfo {author} {Stojchevska L}, \bibinfo {author} {Vaskivskyi I}, \bibinfo
  {author} {Mertelj T}, \bibinfo {author} {Kusar P}, \bibinfo {author} {Svetin
  D}, \bibinfo {author} {Brazovskii S}\ and\ \bibinfo {author} {Mihailovic D},\
  \href {\doibase 10.1126/science.1241591} {\bibfield  {journal} {\bibinfo
  {journal} {Science}\ }\textbf {\bibinfo {volume} {344}},\ \bibinfo {pages}
  {177} (\bibinfo {year} {2014})},\ \Eprint
  {http://arxiv.org/abs/http://science.sciencemag.org/content/344/6180/177.full.pdf}
  {http://science.sciencemag.org/content/344/6180/177.full.pdf} \BibitemShut
  {NoStop}%
\bibitem [{\citenamefont {Gedik}\ \emph
  {et~al.}(2007{\natexlab{a}})\citenamefont {Gedik}, \citenamefont {Yang},
  \citenamefont {Logvenov}, \citenamefont {Bozovic},\ and\ \citenamefont
  {Zewail}}]{Gedik425}%
  \BibitemOpen
  \bibinfo {author} {Gedik N}, \bibinfo {author} {Yang D~S}, \bibinfo {author}
  {Logvenov G}, \bibinfo {author} {Bozovic I}\ and\ \bibinfo {author} {Zewail
  A~H},\ \href {\doibase 10.1126/science.1138834} {\bibfield  {journal}
  {\bibinfo  {journal} {Science}\ }\textbf {\bibinfo {volume} {316}},\ \bibinfo
  {pages} {425} (\bibinfo {year} {2007}{\natexlab{a}})},\ \Eprint
  {http://arxiv.org/abs/http://science.sciencemag.org/content/316/5823/425.full.pdf}
  {http://science.sciencemag.org/content/316/5823/425.full.pdf} \BibitemShut
  {NoStop}%
\bibitem [{\citenamefont {Kim}\ \emph {et~al.}(2012)\citenamefont {Kim},
  \citenamefont {Pashkin}, \citenamefont {Sch?fer}, \citenamefont {Beyer},
  \citenamefont {Porer}, \citenamefont {Wolf}, \citenamefont {Bernhard},
  \citenamefont {Demsar}, \citenamefont {Huber},\ and\ \citenamefont
  {Leitenstorfer}}]{RN211}%
  \BibitemOpen
  \bibinfo {author} {Kim K~W}, \bibinfo {author} {Pashkin A}, \bibinfo {author}
  {Sch?fer H}, \bibinfo {author} {Beyer M}, \bibinfo {author} {Porer M},
  \bibinfo {author} {Wolf T}, \bibinfo {author} {Bernhard C}, \bibinfo {author}
  {Demsar J}, \bibinfo {author} {Huber R}\ and\ \bibinfo {author}
  {Leitenstorfer A},\ \href {\doibase 10.1038/nmat3294} {\bibfield  {journal}
  {\bibinfo  {journal} {Nature Materials}\ }\textbf {\bibinfo {volume} {11}},\
  \bibinfo {pages} {497} (\bibinfo {year} {2012})}\BibitemShut {NoStop}%
\bibitem [{\citenamefont {Fausti}\ \emph {et~al.}(2011)\citenamefont {Fausti},
  \citenamefont {Tobey}, \citenamefont {Dean}, \citenamefont {Kaiser},
  \citenamefont {Dienst}, \citenamefont {Hoffmann}, \citenamefont {Pyon},
  \citenamefont {Takayama}, \citenamefont {Takagi},\ and\ \citenamefont
  {Cavalleri}}]{Fausti189}%
  \BibitemOpen
  \bibinfo {author} {Fausti D}, \bibinfo {author} {Tobey R~I}, \bibinfo
  {author} {Dean N}, \bibinfo {author} {Kaiser S}, \bibinfo {author} {Dienst
  A}, \bibinfo {author} {Hoffmann M~C}, \bibinfo {author} {Pyon S}, \bibinfo
  {author} {Takayama T}, \bibinfo {author} {Takagi H}\ and\ \bibinfo {author}
  {Cavalleri A},\ \href {\doibase 10.1126/science.1197294} {\bibfield
  {journal} {\bibinfo  {journal} {Science}\ }\textbf {\bibinfo {volume}
  {331}},\ \bibinfo {pages} {189} (\bibinfo {year} {2011})}\BibitemShut
  {NoStop}%
\bibitem [{\citenamefont {Kaiser}\ \emph {et~al.}(2014)\citenamefont {Kaiser},
  \citenamefont {Hunt}, \citenamefont {Nicoletti}, \citenamefont {Hu},
  \citenamefont {Gierz}, \citenamefont {Liu}, \citenamefont {{Le Tacon}},
  \citenamefont {Loew}, \citenamefont {Haug}, \citenamefont {Keimer},\ and\
  \citenamefont {Cavalleri}}]{Kaiser2014}%
  \BibitemOpen
  \bibinfo {author} {Kaiser S}, \bibinfo {author} {Hunt C~R}, \bibinfo {author}
  {Nicoletti D}, \bibinfo {author} {Hu W}, \bibinfo {author} {Gierz I},
  \bibinfo {author} {Liu H~Y}, \bibinfo {author} {{Le Tacon} M}, \bibinfo
  {author} {Loew T}, \bibinfo {author} {Haug D}, \bibinfo {author} {Keimer B}\
  and\ \bibinfo {author} {Cavalleri A},\ \href {\doibase
  10.1103/PhysRevB.89.184516} {\bibfield  {journal} {\bibinfo  {journal} {Phys.
  Rev. B}\ }\textbf {\bibinfo {volume} {89}},\ \bibinfo {pages} {184516}
  (\bibinfo {year} {2014})}\BibitemShut {NoStop}%
\bibitem [{\citenamefont {Hu}\ \emph {et~al.}(2014)\citenamefont {Hu},
  \citenamefont {Kaiser}, \citenamefont {Nicoletti}, \citenamefont {Hunt},
  \citenamefont {Gierz}, \citenamefont {Hoffmann}, \citenamefont {Le~Tacon},
  \citenamefont {Loew}, \citenamefont {Keimer},\ and\ \citenamefont
  {Cavalleri}}]{hu2014optically}%
  \BibitemOpen
  \bibinfo {author} {Hu W}, \bibinfo {author} {Kaiser S}, \bibinfo {author}
  {Nicoletti D}, \bibinfo {author} {Hunt C~R}, \bibinfo {author} {Gierz I},
  \bibinfo {author} {Hoffmann M~C}, \bibinfo {author} {Le~Tacon M}, \bibinfo
  {author} {Loew T}, \bibinfo {author} {Keimer B}\ and\ \bibinfo {author}
  {Cavalleri A},\ \href {\doibase 10.1038/nmat3963} {\bibfield  {journal}
  {\bibinfo  {journal} {Nature materials}\ }\textbf {\bibinfo {volume} {13}},\
  \bibinfo {pages} {705} (\bibinfo {year} {2014})}\BibitemShut {NoStop}%
\bibitem [{\citenamefont {Nicoletti}\ \emph {et~al.}(2014)\citenamefont
  {Nicoletti}, \citenamefont {Casandruc}, \citenamefont {Laplace},
  \citenamefont {Khanna}, \citenamefont {Hunt}, \citenamefont {Kaiser},
  \citenamefont {Dhesi}, \citenamefont {Gu}, \citenamefont {Hill},\ and\
  \citenamefont {Cavalleri}}]{Nicoletti2014}%
  \BibitemOpen
  \bibinfo {author} {Nicoletti D}, \bibinfo {author} {Casandruc E}, \bibinfo
  {author} {Laplace Y}, \bibinfo {author} {Khanna V}, \bibinfo {author} {Hunt
  C~R}, \bibinfo {author} {Kaiser S}, \bibinfo {author} {Dhesi S~S}, \bibinfo
  {author} {Gu G~D}, \bibinfo {author} {Hill J~P}\ and\ \bibinfo {author}
  {Cavalleri A},\ \href {\doibase 10.1103/PhysRevB.90.100503} {\bibfield
  {journal} {\bibinfo  {journal} {Phys. Rev. B}\ }\textbf {\bibinfo {volume}
  {90}},\ \bibinfo {pages} {1} (\bibinfo {year} {2014})},\ \Eprint
  {http://arxiv.org/abs/1404.6796} {arXiv:1404.6796} \BibitemShut {NoStop}%
\bibitem [{\citenamefont {Casandruc}\ \emph {et~al.}(2015)\citenamefont
  {Casandruc}, \citenamefont {Nicoletti}, \citenamefont {Rajasekaran},
  \citenamefont {Laplace}, \citenamefont {Khanna}, \citenamefont {Gu},
  \citenamefont {Hill},\ and\ \citenamefont {Cavalleri}}]{PhysRevB.91.174502}%
  \BibitemOpen
  \bibinfo {author} {Casandruc E}, \bibinfo {author} {Nicoletti D}, \bibinfo
  {author} {Rajasekaran S}, \bibinfo {author} {Laplace Y}, \bibinfo {author}
  {Khanna V}, \bibinfo {author} {Gu G~D}, \bibinfo {author} {Hill J~P}\ and\
  \bibinfo {author} {Cavalleri A},\ \href {\doibase 10.1103/PhysRevB.91.174502}
  {\bibfield  {journal} {\bibinfo  {journal} {Phys. Rev. B}\ }\textbf {\bibinfo
  {volume} {91}},\ \bibinfo {pages} {174502} (\bibinfo {year}
  {2015})}\BibitemShut {NoStop}%
\bibitem [{\citenamefont {Zhang}\ \emph
  {et~al.}(2017{\natexlab{a}})\citenamefont {Zhang}, \citenamefont {Wang},
  \citenamefont {Shi}, \citenamefont {Lin}, \citenamefont {Zhang},
  \citenamefont {Gu}, \citenamefont {Dong},\ and\ \citenamefont
  {Wang}}]{Zhang2017b}%
  \BibitemOpen
  \bibinfo {author} {Zhang S~J}, \bibinfo {author} {Wang Z~X}, \bibinfo
  {author} {Shi L~Y}, \bibinfo {author} {Lin T}, \bibinfo {author} {Zhang M~Y},
  \bibinfo {author} {Gu G~D}, \bibinfo {author} {Dong T}\ and\ \bibinfo
  {author} {Wang N~L},\ \href {http://arxiv.org/abs/1712.01174} {\ \textbf
  {\bibinfo {volume} {1}},\ \bibinfo {pages} {1} (\bibinfo {year}
  {2017}{\natexlab{a}})},\ \Eprint {http://arxiv.org/abs/1712.01174}
  {arXiv:1712.01174} \BibitemShut {NoStop}%
\bibitem [{\citenamefont {Singley}\ \emph {et~al.}(2001)\citenamefont
  {Singley}, \citenamefont {Basov}, \citenamefont {Kurahashi}, \citenamefont
  {Uefuji},\ and\ \citenamefont {Yamada}}]{RN90}%
  \BibitemOpen
  \bibinfo {author} {Singley E~J}, \bibinfo {author} {Basov D~N}, \bibinfo
  {author} {Kurahashi K}, \bibinfo {author} {Uefuji T}\ and\ \bibinfo {author}
  {Yamada K},\ \href {\doibase 10.1103/PhysRevB.64.224503} {\bibfield
  {journal} {\bibinfo  {journal} {Physical Review B}\ }\textbf {\bibinfo
  {volume} {64}} (\bibinfo {year} {2001}),\
  10.1103/PhysRevB.64.224503}\BibitemShut {NoStop}%
\bibitem [{\citenamefont {van~der Marel}\ and\ \citenamefont
  {Tsvetkov}(1996)}]{VanderMarel1996}%
  \BibitemOpen
  \bibinfo {author} {van~der Marel D}\ and\ \bibinfo {author} {Tsvetkov A},\
  \href {\doibase 10.1007/BF02548125} {\bibfield  {journal} {\bibinfo
  {journal} {Czechoslovak Journal of Physics}\ }\textbf {\bibinfo {volume}
  {46}},\ \bibinfo {pages} {3165} (\bibinfo {year} {1996})},\ \Eprint
  {http://arxiv.org/abs/9609155} {arXiv:9609155 [cond-mat]} \BibitemShut
  {NoStop}%
\bibitem [{\citenamefont {Shibata}\ and\ \citenamefont
  {Yamada}(1998)}]{Shibata1998}%
  \BibitemOpen
  \bibinfo {author} {Shibata H}\ and\ \bibinfo {author} {Yamada T},\ \href
  {\doibase 10.1103/PhysRevLett.81.3519} {\bibfield  {journal} {\bibinfo
  {journal} {Physical Review Letters}\ }\textbf {\bibinfo {volume} {81}},\
  \bibinfo {pages} {3519} (\bibinfo {year} {1998})}\BibitemShut {NoStop}%
\bibitem [{\citenamefont {Grueninger}\ \emph {et~al.}(1999)\citenamefont
  {Grueninger}, \citenamefont {van~der Marel}, \citenamefont {Tsvetkov},\ and\
  \citenamefont {Erb}}]{Grueninger1999}%
  \BibitemOpen
  \bibinfo {author} {Grueninger M}, \bibinfo {author} {van~der Marel D},
  \bibinfo {author} {Tsvetkov a~a}\ and\ \bibinfo {author} {Erb A},\ \href
  {\doibase 10.1103/PhysRevLett.84.1575} {\bibfield  {journal} {\bibinfo
  {journal} {Physcal Review Letters}\ }\textbf {\bibinfo {volume} {84}},\
  \bibinfo {pages} {7} (\bibinfo {year} {1999})},\ \Eprint
  {http://arxiv.org/abs/9903352} {arXiv:9903352 [cond-mat]} \BibitemShut
  {NoStop}%
\bibitem [{\citenamefont {Timusk}\ and\ \citenamefont
  {Homes}(2003)}]{Timusk2003}%
  \BibitemOpen
  \bibinfo {author} {Timusk T}\ and\ \bibinfo {author} {Homes C~C},\ \href
  {\doibase 10.1016/S0038-1098(02)00666-X} {\bibfield  {journal} {\bibinfo
  {journal} {Solid State Communications}\ }\textbf {\bibinfo {volume} {126}},\
  \bibinfo {pages} {63} (\bibinfo {year} {2003})},\ \Eprint
  {http://arxiv.org/abs/0209371v1} {arXiv:0209371v1 [arXiv:cond-mat]}
  \BibitemShut {NoStop}%
\bibitem [{\citenamefont {Tajima}\ and\ \citenamefont
  {Uchida}(2012)}]{Tajima2012}%
  \BibitemOpen
  \bibinfo {author} {Tajima S}\ and\ \bibinfo {author} {Uchida S~I},\ \href
  {\doibase 10.1016/j.physc.2012.04.002} {\bibfield  {journal} {\bibinfo
  {journal} {Physica C: Superconductivity and its Applications}\ }\textbf
  {\bibinfo {volume} {481}},\ \bibinfo {pages} {55} (\bibinfo {year}
  {2012})}\BibitemShut {NoStop}%
\bibitem [{\citenamefont {Kojima}\ \emph {et~al.}(2002)\citenamefont {Kojima},
  \citenamefont {Uchida}, \citenamefont {Fudamoto},\ and\ \citenamefont
  {Tajima}}]{PhysRevLett.89.247001}%
  \BibitemOpen
  \bibinfo {author} {Kojima K~M}, \bibinfo {author} {Uchida S}, \bibinfo
  {author} {Fudamoto Y}\ and\ \bibinfo {author} {Tajima S},\ \href {\doibase
  10.1103/PhysRevLett.89.247001} {\bibfield  {journal} {\bibinfo  {journal}
  {Phys. Rev. Lett.}\ }\textbf {\bibinfo {volume} {89}},\ \bibinfo {pages}
  {247001} (\bibinfo {year} {2002})}\BibitemShut {NoStop}%
\bibitem [{\citenamefont {LaForge}\ \emph {et~al.}(2007)\citenamefont
  {LaForge}, \citenamefont {Padilla}, \citenamefont {Burch}, \citenamefont
  {Li}, \citenamefont {Dordevic}, \citenamefont {Segawa}, \citenamefont
  {Ando},\ and\ \citenamefont {Basov}}]{PhysRevB.76.054524}%
  \BibitemOpen
  \bibinfo {author} {LaForge A~D}, \bibinfo {author} {Padilla W~J}, \bibinfo
  {author} {Burch K~S}, \bibinfo {author} {Li Z~Q}, \bibinfo {author} {Dordevic
  S~V}, \bibinfo {author} {Segawa K}, \bibinfo {author} {Ando Y}\ and\ \bibinfo
  {author} {Basov D~N},\ \href {\doibase 10.1103/PhysRevB.76.054524} {\bibfield
   {journal} {\bibinfo  {journal} {Phys. Rev. B}\ }\textbf {\bibinfo {volume}
  {76}},\ \bibinfo {pages} {054524} (\bibinfo {year} {2007})}\BibitemShut
  {NoStop}%
\bibitem [{\citenamefont {Rini}\ \emph {et~al.}(2007)\citenamefont {Rini},
  \citenamefont {Tobey}, \citenamefont {Dean}, \citenamefont {Itatani},
  \citenamefont {Tomioka}, \citenamefont {Tokura}, \citenamefont {Schoenlein},\
  and\ \citenamefont {Cavalleri}}]{Rini2007b}%
  \BibitemOpen
  \bibinfo {author} {Rini M}, \bibinfo {author} {Tobey R}, \bibinfo {author}
  {Dean N}, \bibinfo {author} {Itatani J}, \bibinfo {author} {Tomioka Y},
  \bibinfo {author} {Tokura Y}, \bibinfo {author} {Schoenlein R~W}\ and\
  \bibinfo {author} {Cavalleri A},\ \href {\doibase 10.1038/nature06119}
  {\bibfield  {journal} {\bibinfo  {journal} {Nature}\ }\textbf {\bibinfo
  {volume} {449}},\ \bibinfo {pages} {72} (\bibinfo {year} {2007})}\BibitemShut
  {NoStop}%
\bibitem [{\citenamefont {{Vance R. Morrison, Robert. P. Chatelain, Kunal L.
  Tiwari, Ali Hendaoui, Andrew Bruh{\'{a}}cs, Mohamed
  Chaker}}(2014)}]{VanceR2014}%
  \BibitemOpen
  \bibinfo {author} {{Vance R. Morrison, Robert. P. Chatelain, Kunal L. Tiwari,
  Ali Hendaoui, Andrew Bruh{\'{a}}cs, Mohamed Chaker} B~J~S},\ \href {\doibase
  10.1126/science.1253779} {\bibfield  {journal} {\bibinfo  {journal}
  {Science}\ }\textbf {\bibinfo {volume} {346}},\ \bibinfo {pages} {445}
  (\bibinfo {year} {2014})}\BibitemShut {NoStop}%
\bibitem [{\citenamefont {Gedik}\ \emph
  {et~al.}(2007{\natexlab{b}})\citenamefont {Gedik}, \citenamefont {Yang},
  \citenamefont {Logvenov}, \citenamefont {Bozovic},\ and\ \citenamefont
  {Zewail}}]{Gedik2007}%
  \BibitemOpen
  \bibinfo {author} {Gedik N}, \bibinfo {author} {Yang D~s}, \bibinfo {author}
  {Logvenov G}, \bibinfo {author} {Bozovic I}\ and\ \bibinfo {author} {Zewail
  A~H},\ \href {\doibase 10.1126/science.1138834} {\bibfield  {journal}
  {\bibinfo  {journal} {Science}\ }\textbf {\bibinfo {volume} {316}},\ \bibinfo
  {pages} {425} (\bibinfo {year} {2007}{\natexlab{b}})}\BibitemShut {NoStop}%
\bibitem [{\citenamefont {Ichikawa}\ \emph {et~al.}(2011)\citenamefont
  {Ichikawa}, \citenamefont {Nozawa}, \citenamefont {Sato}, \citenamefont
  {Tomita}, \citenamefont {Ichiyanagi}, \citenamefont {Chollet}, \citenamefont
  {Guerin}, \citenamefont {Dean}, \citenamefont {Cavalleri}, \citenamefont
  {Adachi}, \citenamefont {Arima}, \citenamefont {Sawa}, \citenamefont
  {Ogimoto}, \citenamefont {Nakamura}, \citenamefont {Tamaki}, \citenamefont
  {Miyano},\ and\ \citenamefont {Koshihara}}]{Ichikawa2011}%
  \BibitemOpen
  \bibinfo {author} {Ichikawa H}, \bibinfo {author} {Nozawa S}, \bibinfo
  {author} {Sato T}, \bibinfo {author} {Tomita A}, \bibinfo {author}
  {Ichiyanagi K}, \bibinfo {author} {Chollet M}, \bibinfo {author} {Guerin L},
  \bibinfo {author} {Dean N}, \bibinfo {author} {Cavalleri A}, \bibinfo
  {author} {Adachi S~i}, \bibinfo {author} {Arima T~h}, \bibinfo {author} {Sawa
  H}, \bibinfo {author} {Ogimoto Y}, \bibinfo {author} {Nakamura M}, \bibinfo
  {author} {Tamaki R}, \bibinfo {author} {Miyano K}\ and\ \bibinfo {author}
  {Koshihara S~y},\ \href {\doibase 10.1038/NMAT2929} {\bibfield  {journal}
  {\bibinfo  {journal} {Nature materials}\ }\textbf {\bibinfo {volume} {10}},\
  \bibinfo {pages} {101} (\bibinfo {year} {2011})}\BibitemShut {NoStop}%
\bibitem [{\citenamefont {Mankowsky}\ \emph {et~al.}(2014)\citenamefont
  {Mankowsky}, \citenamefont {Subedi}, \citenamefont {Forst}, \citenamefont
  {Mariager}, \citenamefont {Chollet}, \citenamefont {Lemke}, \citenamefont
  {Robinson}, \citenamefont {Glownia}, \citenamefont {Minitti}, \citenamefont
  {Frano}, \citenamefont {Fechner}, \citenamefont {Spaldin}, \citenamefont
  {Loew}, \citenamefont {Keimer}, \citenamefont {Georges},\ and\ \citenamefont
  {Cavalleri}}]{RN138}%
  \BibitemOpen
  \bibinfo {author} {Mankowsky R}, \bibinfo {author} {Subedi A}, \bibinfo
  {author} {Forst M}, \bibinfo {author} {Mariager S~O}, \bibinfo {author}
  {Chollet M}, \bibinfo {author} {Lemke H~T}, \bibinfo {author} {Robinson J~S},
  \bibinfo {author} {Glownia J~M}, \bibinfo {author} {Minitti M~P}, \bibinfo
  {author} {Frano A}, \bibinfo {author} {Fechner M}, \bibinfo {author} {Spaldin
  N~A}, \bibinfo {author} {Loew T}, \bibinfo {author} {Keimer B}, \bibinfo
  {author} {Georges A}\ and\ \bibinfo {author} {Cavalleri A},\ \href {\doibase
  10.1038/nature13875} {\bibfield  {journal} {\bibinfo  {journal} {Nature}\
  }\textbf {\bibinfo {volume} {516}},\ \bibinfo {pages} {71} (\bibinfo {year}
  {2014})}\BibitemShut {NoStop}%
\bibitem [{\citenamefont {Li}\ \emph {et~al.}(2008)\citenamefont {Li},
  \citenamefont {Chi}, \citenamefont {Zhao}, \citenamefont {Wen}, \citenamefont
  {Stone}, \citenamefont {Lynn},\ and\ \citenamefont {Dai}}]{Li2008}%
  \BibitemOpen
  \bibinfo {author} {Li S}, \bibinfo {author} {Chi S}, \bibinfo {author} {Zhao
  J}, \bibinfo {author} {Wen H~H}, \bibinfo {author} {Stone M~B}, \bibinfo
  {author} {Lynn J~W}\ and\ \bibinfo {author} {Dai P},\ \href {\doibase
  10.1103/PhysRevB.78.014520} {\bibfield  {journal} {\bibinfo  {journal} {Phys.
  Rev. B}\ }\textbf {\bibinfo {volume} {78}},\ \bibinfo {pages} {17} (\bibinfo
  {year} {2008})},\ \Eprint {http://arxiv.org/abs/arXiv:0806.3426v1}
  {arXiv:arXiv:0806.3426v1} \BibitemShut {NoStop}%
\bibitem [{\citenamefont {Zhang}\ \emph
  {et~al.}(2017{\natexlab{b}})\citenamefont {Zhang}, \citenamefont {Wang},
  \citenamefont {Dong},\ and\ \citenamefont {Wang}}]{Zhang2017}%
  \BibitemOpen
  \bibinfo {author} {Zhang S~J}, \bibinfo {author} {Wang Z~X}, \bibinfo
  {author} {Dong T}\ and\ \bibinfo {author} {Wang N~L},\ \href {\doibase
  10.1007/s11467-017-0716-4} {\bibfield  {journal} {\bibinfo  {journal}
  {Frontiers of Physics}\ }\textbf {\bibinfo {volume} {12}},\ \bibinfo {pages}
  {127802} (\bibinfo {year} {2017}{\natexlab{b}})},\ \Eprint
  {http://arxiv.org/abs/1708.01991} {arXiv:1708.01991} \BibitemShut {NoStop}%
\end{thebibliography}

%

\begin{small}
 {\textbf{Acknowledgement}}\par
This work was supported by the National Science Foundation of China (No. 11327806, GZ1123) and the National Key Research and Development Program of China (No.2016YFA0300902, 2017YFA0302903, 2017YFA0302904). P. D. is supported by the U.S. DOE, BES under Contract No. DE-SC0012311 and the Robert A. Welch Foundation Grant No. C-1839.

{\textbf{Author contribution}}\par
This project was initiated and supervised by N. L. Wang. Samples was provided by S. L. Li and P. C. Dai. T. Dong, S. J. Zhang and Z. X. Wang built the tunable NIR to MIR pump-THz probe spectroscopy system. S. J. Zhang and D. Wu performed the optical measurements. The experimental data was analysed by N. L. Wang, S. J. Zhang and Z. X. Wang. The manuscript was written by N. L. Wang and S. J. Zhang with input from all authors.

{\textbf{Competing financial interests}}\par
The authors declare no competing financial interests.

\end{small}

\end{document}